\documentstyle{plenum}

\def\ov{\over}
\title{TOWARDS A FIELD THEORETICAL DESCRIPTION OF\\
MULTIPARTICLE PRODUCTION IN HIGH ENERGY\\
COLLISIONS\footnote{\rm\baselineskip=11pt
Presented by I. Sarcevic at {\it NATO Advanced Study Workshop on
Hot Hadronic Matter: Theory and Experiment}, Divonne-les-Bains,
Switzerland, June 27 - July 1, 1994.}}
\author{Ina Sarcevic${^{1}}$
and H. Th. Elze${^{2}}$\\ \\
${^{1}}$Department of Physics\\
\ \ University of Arizona, Tucson, AZ85721, USA \\
${^{2}}$Theory Division, CERN\\
\ \ CH-1211 Geneva 23, Switzerland}
\begin{document}
\maketitle                 
\vglue 0.8cm
\begin{center}
{\bf ABSTRACT}
\end{center}
\vglue 0.3cm
{\rightskip=3pc
 \leftskip=3pc
 \rm\baselineskip=12pt
 \noindent
We present an effective field theory of
multiparticle correlations based on analogy with Ginzburg-Landau
theory of superconductivity.  With the assumption
that the field
represents
particle density fluctuations, and
in the case of
gaussian-type
effective action
we find that there are
no higher-order
correlations, in agreement with the recent observations in
high energy heavy-ion
collisions.
We predict that
three-dimensional
two-particle correlations have Yukawa form.  We also present our
results for the
two-dimensional and one-dimensional
two-particle correlations (i.e. cumulants)
as projections of our theory to lower dimensions.
}

\bigskip

\noindent{\bf{1.  INTRODUCTION}}

\medskip

Unusually large fluctuations in
transverse energy
have recently been observed
in ultrarelativistic heavy-ion collisions
at CERN energies.$^1$
The
independent-collision model which contains the
scatterings of the secondary nucleons fails to describe the observed
data indicating
that perhaps nuclear constituents scatter and
produce particle coherently.$^2$
Similar conclusion has been reached when
multiparticle density
fluctuations in different
phase space regions have been studied
via
factorial moments, defined as
$$F_{q}(\delta y) = {1\over M} \sum_{m=1}^{M} {\langle n_{m}(n_{m}-1)
\ldots (n_{m}-i+1)\rangle \over  \langle n_{m} \rangle^{i}},\eqno(1)$$
where M is
the number of rapidity bins $(\delta y = Y/M )$ and $n_m$ is the
number of particles in the $m^{th}$ bin.$^3$
In the past few years,
these moments have
been measured for different targets and
projectiles
at energy of
$200$GeV/nucleon
.$^4$  They
were found to
increase
with decreasing bin size
indicating nonstatistical fluctuations and being
incompatible with
the predictions of the
standard
classical hadronization models embedded in the existing Monte Carlo
models.$^4$  In addition,
the observed effect, sometimes referred to as the
``intermittency effect'', can not be accounted for by the
superposition of independent nucleon-nucleon collisions, even
when rescattering and geometrical effects are included.$^3$

The possibility of creating
the new form of matter, the quark-gluon plasma, in high-energy
heavy-ion collisions
have inspired intensive
theoretical work on identifying
the unambiguous QGP signal.
Thus it is not surprising that the
observation of the unusually large
multiparticle density
fluctuations has created a new excitement in the field,
especially as a possibility of
pointing towards the
onset of the
phase transition from quark-gluon plasma to hadronic matter.
Phase transitions in QCD at high
temperatures are of general interest -- they are
directly relevant to cosmology, since such a phase transition
occurred throughout the universe during the
early moments of the big bang and
a first order phase
transition could have
altered primordial nuclear abundances.
Unfortunately,
up to now
there
are no conclusive predictions for detecting the
quark
matter in heavy-ion collisions
and there is no
theory
to describe the observed ``intermittency'' phenomenon.$^5$

\bigskip
\noindent{\bf {2.  MULTIPARTICLE CORRELATIONS IN HIGH-ENERGY
COLLISIONS}}
\medskip

Multiparticle correlations in
three ``dimensions'' are usually measured by subdividing
a given total interval
$\Omega_{\rm tot} = \Delta Y\, \Delta \phi\, \Delta P$
into $M^3$ bins of side lengths \ \ \ \ \ \ \ \

\noindent
$(\Delta Y/M,\, \Delta \phi/M,\, \Delta P/M)$. With $n_{klm}$
the number of particles in bin $(k,l,m)$
and $n^{[q]} = n!/(n-q)!$
the ``vertical'' factorial moment is

$$F_q^v(M) \equiv {1\over M^3}
\sum_{k,l,m=1}^M
{
\langle n_{klm}^{[q]} \rangle
\over
\langle n_{klm} \rangle^{q}
}
=
{1\over M^3}
\sum_{k,l,m=1}^M
{
\int_{\Omega_{klm}} \prod_i  d^3 x_i \, \rho_q(\vec{x}_1\ldots
\vec{x}_q)
\over
\left[\int_{\Omega_{klm}} d^3 x \, \rho_1(\vec{x})\right]^q
}. \eqno(2)$$
\noindent
The second equality illustrates how the factorial moment can be
written in terms of integrals of the correlation function $\rho_q$
($\Omega_{klm}$ is the region of integration over bin $k,l,m$).$^6$
  Because for small bin sizes $n_{klm}$ becomes small and the relative
error correspondingly large, an alternative definition is often
preferred for three-dimensional analysis, the ``horizontal''
factorial moment,
$$
F_q^h(M) \equiv {1\over M^3}
\sum_{k,l,m=1}^M
{
\langle n_{klm}^{[q]} \rangle
\over
\left(\langle N \rangle/M^3\right)^{q}
}  \nonumber\\
=
M^{3(q-1)}
\sum_{k,l,m=1}^M
{
\int_{\Omega_{klm}} \prod_i  d^3 x_i \, \rho_q(\vec{x}_1\ldots
\vec{x}_q)
\over
\left[\int_{\Omega_{\rm tot}} d^3 x \, \rho_1(\vec{x})\right]^q
} .
\eqno(3)$$

This form, while being much more stable,
has the drawback that it depends
on the shape of the one-particle distribution function
$\rho_1$.

In order to examine the true higher-order correlations, the
trivial, combinatoric contributions
from two-particle correlations need to
be subtracted.
The cumulant moments, $K_q$, which measure the true, dynamical
correlations are defined as$^7$

$$K_q^v (\Omega_m) =
{1 \ov M^3} \sum_m \int_{\Omega_m} \prod_i d^3\vec x_i
{}~~~
{k_2(\vec x_1,\vec x_2\ldots \vec x_q) \rho_1(\vec x_1) \rho_1(\vec x_2)
\over
\left[\int_{\Omega_{klm}} d^3 \vec x\, \rho_1(\vec x)
\right]^q}
,\eqno(4)$$
where
$$k_2(1,2)= {\rho_2(\vec{x}_1,\vec{x}_2)\over {<\rho
(\vec{x}_1)><\rho(\vec{x}_2)>}}-1, \eqno(5)$$
$$k_3(1,2,3)={\rho_3(\vec{x}_1,
\vec{x}_2,\vec{x}_3)\over {<\rho(\vec{x}_1)><\rho(\vec{x}_2)>
<\rho(\vec{x}_3)>}}
-\sum_{perm}^{(3)} {\rho_2(\vec{x}_1,\vec{x}_2)\over
{<\rho(\vec{x}_1)><\rho(\vec{x}_2)>}}
+ 2 ,$$
$$k_4=
{\rho_4(\vec{x}_1,\vec{x}_2,\vec{x}_3,\vec{x}_4))
\over {<\rho(\vec{x}_1)><\rho(\vec{x}_2)><\rho(\vec{x}_3)>
<\rho(\vec{x}_4)>}}
-\sum_{perm}^{(4)} {\rho_3(\vec{x}_1,\vec{x}_2,\vec{x}_3))
\over {<\rho(\vec{x}_1)><\rho(\vec{x}_2)><\rho(\vec{x}_3)>}}
$$
$$-\sum_{perm}^{(3)} {\rho_2(\vec{x}_1,\vec{x}_2)\rho_2
(\vec{x}_3,\vec{x}_4))
\over {<\rho(\vec{x}_1)><\rho(\vec{x}_2)><\rho(\vec{x}_3)>
<\rho(\vec{x}_4)>}}
+\sum_{perm}^{(12)} {\rho_2(\vec{x}_1,\vec{x}_2)
\over {<\rho(\vec{x}_1)><\rho(\vec{x}_2)>}}
-6.\eqno(6)$$

The factorial moments,
$F_q$ ($q=3,4,5$), given by Eq. (2)
can be expressed in terms of the cumulans in the following way:$^8$

$$F_2 = 1+ K_2,
$$
$$F_3 = 1+ 3K_2 +K_3, $$
$$F_4 = 1+ 6K_2 + 3 {(K_2)^2} + 4K_3 + K_4, \eqno(7)$$
$$F_5 = 1+ 10K_2 + 15 {(K_2)^2} + 10 {K_3 K_2}
     + 10K_3 + 5 K_4 + K_5.$$
Clearly, if there are no true, dynamical correlations, the
cumulants, $K_q$ vanish and factorial moments approach unity.

It has been found that
$K_2$ decreases from lighter to heavier projectiles
, especially in the case of Sulfur.$^9$
Furthermore, in hadronic collisions $K_3$ and $K_4$ are
non-negligible (for example, $K_3$
contributes up to $20\%$ to $F_3$ at small
$\delta y$), while
in nucleus-nucleus collisions, at the same energy,
these cumulants are compatible with zero.$^8$
This implies
that there are no statistically significant
correlations of order higher than two for heavy-ion
collisions and that
the observed increase of the higher-order factorial
moments $F_q$ is entirely due to the
dynamical two-particle correlations.  As an illustration,
in Fig. 1 we present the
cumulant $K_3(M)$ from
the NA35 data for O-Au at $200$GeV/nucleon.
Recent NA35 two and three dimensional measurements of the
factorial moments corroborate our findings.$^{9}$

\vskip 3.50 true in
\noindent
Figure 1:
Third order cumulant $K_3$ as a function of the number
of bins $M$ for NA35 OAu data in $(y,\phi,p_\perp)$.$^{9}$
Cumulants of higher order are also compatible with zero. This is
confirmed in analyses in terms of other variables and different
colliding nuclei.$^{10}$
\vfil\eject

\bigskip
\noindent{\bf {3.  EFFECTIVE FIELD THEORY OF MULTIPARTICLE PRODUCTION}}
\medskip

The fact that particles
produced in high-energy heavy-ion collisions exhibit
only two-particle correlations indicates that perhaps
higher-order correlations are washed out by rescattering
of the initially correlated particles.  Presently, there is
 no theory that describes this phenomena.
Recently, we have proposed
a three-dimensional
statistical field theory of density
fluctuations which has these features.$^{11,12}$
This model was formulated in analogy with the Ginzburg-Landau
theory of superconductivity.  The large
number of particles produced in ultrarelativistic
heavy-ion collisions justifies the use if a
statistical theory of particle production.
The formal analogy with the statistical mechanics
of a one-dimensional ``gas'' was first pointed out by
Feynman and Wilson and was
later further developed by
Scalapino and Sugar$^{13}$ and many others.$^5$
The idea is to build a
statistical theory of the macroscopic observables by
imagining that the microscopic degrees of freedom are
integrated out and represented in terms of a few
phenomenological parameters
and by postulating that
this theory
will
eventually be derived from a more
fundamental theory, such
as QCD.
\par
While in the G-L theory of superconductivity
the field (i.e. the order parameter) represents superconducting
pairs, in the particle production problem, the relevant
variable is the density fluctuation.
The ``field'' $\Phi(\vec{x})$ is
a random variable which depends on the rapidity of the
particle and
its transverse momentum $p_t$ and it is
identified with the density fluctuation.
\par
Even though particles
produced in high-energy collisions need not be in
thermal equilibrium, one can still introduce a
functional of the field $\Phi$, $F[\Phi]$, which plays
a role analogous to the free energy in equilibrium
statistical mechanics.  In principle one should be
able to derive this functional from the underlying
dynamics.
\par
We define a random field $\Phi$ as a function in a
three-dimensional space spanned by $(y,\phi,p_\perp)$.
Throughout, $p_\perp$ will be implicitly divided by a constant
scale $\cal P$ so that it is dimensionless. Since we are not
looking for a phase transition, we omit the quartic term
and start with the functional$^{11}$
$$
F[\Phi] = \int_0^P dy\, \int_{-P/2}^{P/2} d^2p_\perp
\left[ a^2 \left(\partial\Phi / \partial y\right)^2
     + a^2 \left(\nabla_{\vec p_\perp} \Phi \right)^2
     + \mu^2 \Phi^2
\right] \;. \eqno(8)$$
Taking the appropriate functional derivative, we find for the functional
(8) the three-dimensional form of the two-point function
$$
\langle \Phi(\vec x_1)\Phi(\vec x_2) \rangle
= {1\over 8 \pi a^2} {e^{-R/\xi} \over R} \;,
\eqno(9)$$
where $\xi = a/\mu$ and
$
R  \equiv
[(y_1-y_2)^2 + p_{\perp 1}^2 + p_{\perp 2}^2
- 2p_{\perp 1} p_{\perp 2}\cos(\phi_1-\phi_2) ]^{1/2}
$.
Further, we
define $\Phi(\vec x)$ as the fluctuation at the point $\vec x$
of the particle density for a given event, $\hat\rho_1(\vec x)$,
above/below the mean single particle distribution $\rho_1$ at that
point:
$$
\Phi(\vec x) \equiv {\hat\rho_1(\vec x) \over \rho_1(\vec x) } - 1 \;.
\eqno(10)$$
Through these definitions, we find that
$
\langle \Phi(\vec x_1)\Phi(\vec x_2) \rangle \;
= \;  k_2(\vec x_1,\vec x_2)
$
and that all higher order cumulants become exactly
zero,  $k_{q\ge 3} = 0$.
By means of the specific form of the
functional (8) and the definition of $\Phi$ as a
fluctuation, we take account of the experimental facts in this
regard. What is not experimentally certain and is to be tested is
whether the second order correlations obey the Yukawa form
(9).
\vfil\eject
\smallskip
\noindent{\bf 4. PROJECTIONS OF MULTIPARTICLE CORRELATIONS
(i.e. CUMULANTS) TO LOWER DIMENSIONS}

\smallskip

The second reduced cumulant $k_2 \propto e^{-R/\xi}/R$ can
be compared to data only after a suitable integration over
its variables. For three dimensions, the vertical
integrated cumulant is given by
$K_2^v(\delta y,\delta\phi,\delta p)
=  F_2^v - 1
=  M^{-3}\sum_{k,l,m=1}^M K_2^v(k,l,m) \;,
$
(always taking $\vec x \equiv (y,\phi,p_\perp)$), with
$$
K_2^v(k,l,m)    =
{
\int_{\Omega_{klm}}d^3\vec x_1\, d^3 \vec x_2\; C_2(\vec x_1,\vec x_2)
\over
\left[
\int_{\Omega_{klm}}d^3 \vec x\; \rho_1(\vec x)
\right]^2
}
=
\int_{\Omega_{klm}} d^3\vec x_1\, d^3\vec x_2
{
k_2(\vec x_1,\vec x_2) \rho_1(\vec x_1) \rho_1(\vec x_2)
\over
\left[\int_{\Omega_{klm}} d^3 \vec x\, \rho_1(\vec x)
\right]^2
} \;,
\eqno(11)$$
i.e.\ the integration of $k_2$ involves a correction due to the shape of
the one-particle three-dimensional distribution function
$\rho_1(\vec x)$. Eq.\ (11) as it stands is exact;
horizontal versions have also been derived.$^{12}$
A first test
of our model would therefore be to see if Eq. (7) or its
horizontal equivalent obeys the data in $(y,\phi,p_\perp)$.

The theoretical $k_2(\vec x_1,\vec x_2)$ is further tested by
comparing to factorial cumulant data of lower dimensions.
For example, in $(y,\phi)$, the cumulant is
$$K_2^v(\delta y,\delta\phi) = M^{-2} \sum_{lm} K_2^v(l,m)
\eqno(12)$$
with
$p_\perp$ integrated over the whole window $\Delta P$,
$$
K_2^v(l,m) =
\int_{\Omega_m}dy_1 dy_2 \int_{\Omega_l}d\phi_1 d\phi_2
\int_{\Delta P} dp_1 dp_2\,
{
 k_2(\vec x_1,\vec x_2)\, \rho_1(\vec x_1)\rho_1(\vec x_2)
\over
\left[ \int_{\Omega_m}dy \int_{\Omega_l}d\phi \int_{\Delta P} dp\;
\rho_1(\vec x) \right]^2
} .
\eqno(13)$$
Cumulants of other variable combinations and lower dimensions
are obtained analogously.
With these relations it is thus possible, given any
three-dimensional theoretical function $k_2$ to
compute factorial cumulants and moments for any combination of
its variables. Doing this for different variables serves as
a strong test of the theoretical function as the moments probe
its different regions.

In Figures 2-3
we present our results for the vertical and
horizontal factorial moments.
In our calculation of the projections, we make the
following approximations:
We factorize
the one-particle distribution into its separate variables:
$\rho_1(\vec x) = \langle N \rangle_\Omega\,
                  g(y)\, h(\phi)\, f(p_\perp)$,
where the three distributions $g$, $h$ and $f$ are separately normalized
over their respective total intervals $\Delta Y$, $\Delta \Phi$
and $\Delta P$.
The azimuthal distribution is taken as flat, $h(\phi) = 1/\Delta\Phi$.
We use the full experimental parametrization for $f(p_\perp)$
provided by NA35.
The choice of
two parameters $a$ and $\xi$ in Figures 2-3
were given
only as an illustration.
They need to be determined by comparison with
three-dimensional data.
Once they are fixed, the two-dimensional and one-dimensional
projections are genuine predictions of the theory.

\bigskip
\noindent{\bf {5.  SUMMARY}}

\medskip

In summary, we have presented a three-dimensional
effective
field theory of multiparticle correlations, which gives
no higer-order
correlations, in agreement with the recent heavy-ion data.
In our theory, two-particle
correlations have Yukawa form and the corresponding integrated
cumulants have singular behavior for small regions of phase space.
This prediction seems to be in qualitative agreement with the
recent
NA35 data.$^{14}$
In addition, we have shown that once the parameters $a$ and $\xi$ are
determined from comparison with three-dimensional data,
our theory gives genuine predictions
for the two-dimensional and one-dimensional cumulants.
It will be interesting to see whether all
our predictions are
confirmed with future heavy-ion data.
\vfil\eject
\centerline{ }
\noindent
\vskip 4.0true in
\noindent
Figure 2:
Theoretical predictions for the
vertical cumulant moments $K_2^v$ for
various dimensions, for fixed parameters $a=2.0$, $\xi = 1.0$,
incorporating the experimental NA35 rapidity and $p_\perp$
distributions for 200 A GeV O+Au.
\vskip 4.5 true in
\noindent
Figure 3:
Theoretical predictions for the
horizontal cumulant moments
$K_2^h$ for the same fixed parameters and NA35 distributions
as in Fig.\ 2. The effect of the $p_\perp$ distribution is apparent.
Comparison with NA35 data requires
conversion to horizontal factorial moments $F_2^h$ and a fit
of $a$ and $\xi$.
\vfil\eject
\smallskip
\subsection*{Acknowledgements}
\smallskip
Part of the work presented here
was done in collaboration with H. Eggers whom we thank
for many interesting discussions.  This work was
supported in part by the
DOE grants DE-FG03-93ER40792 and
DE-FG02-85ER40213.

\bigskip
\noindent{\bf {REFERENCES}}
\smallskip

\begin{itemize}
\item{1.}  NA34 Collaboration, F. Corriveau {\it et al.},
{\it Z. Phys.} {\bf C38}, 15 (1988); NA34 Collaboration,
J. Schukraft {\it et al.} {\it Z. Phys.} {\bf C38},
59 (1988).

\item{2.}  G. Baym, G. Friedman and I. Sarcevic, {\it
Phys. Lett.} {\bf B219}, 205 (1989).

\item{3.}  A. Capella, K. Fialkowski and A. Krzywicki,
{\it Phys. Lett.} {\bf B230}, 149 (1989).

\item{4.}  For a recent experimental review, see W. Kittel,
in {\it
Proceedings of the XXIII International Symposium on Multiparticle
Dynamics}, eds. M. Block and A. White (World Scientific,
Singapore, 1994), pg. 251.

\item{5.}
For a recent theoretical review, see R. Hwa,
in {\it
Proceedings of the XXIII International Symposium on Multiparticle
Dynamics}, eds. M. Block and A. White (World Scientific,
Singapore, 1994), pg. 239.

\item{6.}
P. Carruthers and I. Sarcevic, {\it Phys. Rev. Lett.}
{\bf 63}, 1562 (1989).

\item{7.}
P.\ Carruthers, H.C.\ Eggers, and I.\ Sarcevic,
{\it Phys. Lett.} {\bf 254B}, 258 (1991).

\item{8.}
P.\ Carruthers, H.C.\ Eggers
and I.Sarcevic,
{\it  Phys.\ Rev.} {\bf C44}, 1629 (1991).

\item{9.}  NA35 Collaboration, I.\ Derado, in
{\it Proceedings of the
Ringberg Workshop on
Multiparticle Production},
ed.\ R.C.\ Hwa, W.\ Ochs and N.\ Schmitz
(World Scientific, 1992) pg. 184;
NA35 Collaboration, P.\ Seyboth {\it et al.}
, {\it Nucl. Phys.} {\bf A544}
293 (1992).

\item{10.}  H. C. Eggers, Ph. D. Thesis, University of Arizona, 1991.

\item{11.}  H.-Th.\ Elze and I.\ Sarcevic,
{\it Phys. Rev. Lett.}  {\bf 68}, 1988 (1992).

\item{12.}  H.C.\ Eggers, H.T.\ Elze and I.\ Sarcevic,
{\it Int. J. Mod. Phys.} {\bf A9}, 3821 (1994).

\item{13.}
D. J. Scalapino and R.\ L.\ Sugar,
               {\it Phys.\ Rev.} {\bf D8}, 2284 (1973);
             J.C.\ Botke, D.J.
Scalapino and R.L.\ Sugar,
{\it ibid.}, {\bf D9}, 813 (1974);
{\it ibid.}, {\bf 10}, 1604 (1974).

\item{14.}  NA35 Collaboration, P. Seyboth,
in {\it
Proceedings of the XXIII International Symposium on Multiparticle
Dynamics}, eds. M. Block and A. White (World Scientific,
Singapore, 1994), pg. 325.

\end{itemize}

\end{document}